\documentclass[conference]{IEEEtran}
\usepackage{cite}
\usepackage{amsmath,amssymb,amsfonts}
\usepackage{graphicx}
\usepackage{booktabs}
\usepackage{siunitx}
\usepackage{xcolor}
\usepackage{tikz}
\usetikzlibrary{arrows.meta,positioning,fit,calc}
\usepackage{pgfplots}
\pgfplotsset{compat=1.18}
\usepgfplotslibrary{groupplots}
\usepackage[caption=false,font=footnotesize]{subfig}
\usepackage[hidelinks]{hyperref}
\graphicspath{{figs/}}
\definecolor{cEight}{HTML}{1F5FBF}
\newcommand{\eo}[1]{{\color{cEight}#1}}
\newenvironment{eoblock}{\par\begingroup\color{cEight}}{\endgroup\par}

\newcommand{\fD}{f_{\mathrm{D}}}
\newcommand{\vS}{\mathbf{v}_{\mathrm{S}}}

\definecolor{cT0}{HTML}{2A6FBD}\definecolor{cT1}{HTML}{2A9D8F}\definecolor{cT2}{HTML}{7B3FA0}\definecolor{cT3}{HTML}{C97A1E}\definecolor{cT4}{HTML}{B3261E}
\definecolor{cQ}{HTML}{2A6FBD}\definecolor{cS}{HTML}{C97A1E}\definecolor{cH}{HTML}{7B3FA0}\definecolor{cWall}{HTML}{B3261E}
\pgfplotsset{
  resaxis/.style={width=\columnwidth, height=4.3cm, font=\scriptsize, grid=major, grid style={gray!25,very thin},
    axis line style={gray!60}, tick style={gray!60}, xmin=-0.3, xmax=14, enlarge y limits=false,
    every axis plot/.append style={mark size=1.25pt, line width=0.5pt}},
}

\begin{document}

\title{Analysis of the Effect of Doppler on Direct-to-Cell: Real-World Performance and Proposed Improvements}

\author{\IEEEauthorblockN{Yuxi Chen, Muhammad Iqbal Rochman, Kyler McCormack, Jakob Kotz, and Monisha Ghosh}
\IEEEauthorblockA{Department of Electrical Engineering, University of Notre Dame, Notre Dame, IN, USA\\
ychen67@nd.edu, mrochman@nd.edu, kmccorm7@nd.edu, jkotz@nd.edu, mghosh3@nd.edu}
}

\maketitle

\begin{abstract}
Direct-to-cell (DtC) satellites serve unmodified LTE phones today from low Earth orbit, where the satellite's motion creates Doppler shifts of tens of kHz. Since large offsets such as these cannot be compensated adequately by the user equipment (UE), the satellite pre-compensates the Doppler shift of each beam, but it can do so for only one reference point in the beam. A UE elsewhere in the beam still has to compensate a residual offset; previous field studies report residuals of up to \SI{10}{kHz}, beyond the $\pm$\SI{7.5}{kHz} that the \SI{15}{kHz} subcarrier spacing can resolve. In this paper we corroborate this effect with our own detailed measurements of Starlink transmissions and further correlate ineffective handover performance to inadequate residual offset compensation. We analyze the Doppler compensation problem with detailed standard-specific simulations and propose a compensation method at the UE, based on the location of the UE with respect to the beam reference point used for the Doppler compensation and the known satellite location, that improves the efficiency of handover. With UE cell-specific pre-compensation, the neighbors are identified, read correctly and handed over to in time rather than lost to radio-link failure.
\end{abstract}


\section{Introduction}
\label{sec:intro}
Direct-to-cell (DtC) satellites in low Earth orbit (LEO) serve unmodified LTE user equipment (UE). In the US, one \SI{5}{MHz} carrier with \SI{15}{kHz} subcarrier spacing (SCS) is used in the PCS G block, LTE band~25 (downlink 1990--\SI{1995}{MHz}, uplink 1910--\SI{1915}{MHz}) \cite{space_exploration_technologies_corp_spacex_sched_2025}, and each beam is one LTE cell \cite{noauthor_spacex_nodate}. At the \SI{525}{km} and \SI{530}{km} altitudes of the two filed orbits (inclination $53^\circ$ and $43^\circ$) \cite{space_exploration_technologies_corp_spacex_sched_2025}, the Doppler shift spans about $\pm$\SI{40}{kHz} over a pass. The satellite pre-compensates each cell for the Doppler at one Earth-fixed reference point, presumably the beam center \cite{noauthor_spacex_nodate,liu_variegated_2026}. A UE elsewhere in the cell keeps a residual $r$: its own Doppler minus that of the reference point.

On the live US service (Section~\ref{sec:field}), phones locked to the DtC carrier (EARFCN 8665) change cell mostly through radio-link failures (RLFs), not handovers. Before an RLF, the serving reference signal received power (RSRP) falls \SI{13}{dB} below its own level, against \SI{4}{dB} before a terrestrial handover, and the downlink modulation and coding scheme (MCS) falls from about 11 to 4. This paper asks why the UE does not hand over in time, and how this can be fixed. It argues that the cause is the neighbor measurement: the UE is settled on its serving cell, but measures each neighbor at that serving cell's carrier. As a result, the neighbor is missed or read low (Section~\ref{sec:background}). 

To this end, the paper (i)~shows from field data that cell changes on DtC are mostly late RLFs, with the serving cell far below its level and a much stronger new cell (Section~\ref{sec:field}); (ii)~simulates a legacy UE at link level, to find how far from the reference point it can be before its RSRP measurement breaks down, for several geometries, and how its neighbor report fails as the serving residual grows (Sections~\ref{sec:sim} and~\ref{sec:results}); and (iii)~proposes UE pre-compensation per beam, which removes each cell’s predicted residual and restores the measurement, but needs the reference points of both cells and satellites' locations (Section~\ref{sec:discussion}).

\subsubsection*{State of the Art}
Residual-Doppler studies address the NB-IoT uplink \cite{kim_performance_2022}, NR downlink initial synchronization within $\pm$SCS/2 \cite{meshram_doppler_2023} and NR uplink pre-compensation \cite{lin_doppler_2021}. Mozo et al.\ \cite{mozo_synchronization_nodate} and the 3GPP NTN specifications \cite{noauthor_tr_nodate,noauthor_ts_nodate-2} target NTN-capable UEs. DtC measurement studies give no physical-layer model \cite{liu_variegated_2026,garcia-cabeza_direct--cell_2026}. None of them examines the mechanism hypothesized here for DtC LTE or checks it against field data.

\section{Background and Hypothesis}
\label{sec:background}
\subsection{The \SI{15}{kHz} Wall}
LTE band~25 uses a subcarrier spacing $\Delta f_{\mathrm{sc}}=\SI{15}{kHz}$. A frequency offset $\Delta f$ moves every subcarrier off the receiver's fast Fourier transform (FFT) grid. Each grid point then keeps only $\mathrm{sinc}^2(\Delta f/\Delta f_{\mathrm{sc}})$ of its own subcarrier; the rest leaks to the others as inter-carrier interference (ICI)\eo{ (Fig.~\ref{fig:wall}b)}. The cyclic prefix (CP) is a copy of the last samples of each symbol, placed at its start. A receiver that estimates the offset from the CP compares the CP with the samples it copies, $T_u=1/\Delta f_{\mathrm{sc}}$ later. Their phase difference, $2\pi\Delta f/\Delta f_{\mathrm{sc}}$, is known only modulo $2\pi$ \cite{van_de_beek_ml_1997}. The CP estimator therefore sees only the fractional part of the offset, $|\Delta f|<\Delta f_{\mathrm{sc}}/2=\SI{7.5}{kHz}$, the CP reach. For example, a residual of \SI{10}{kHz} is estimated as \SI{-5}{kHz}. After the correction, the offset is \SI{15}{kHz}\eo{ (Fig.~\ref{fig:wall}c--d)}, and each resource element carries the content of its neighbor.

\begin{eoblock}
\begin{figure}[t]
\centering
\includegraphics[width=\columnwidth]{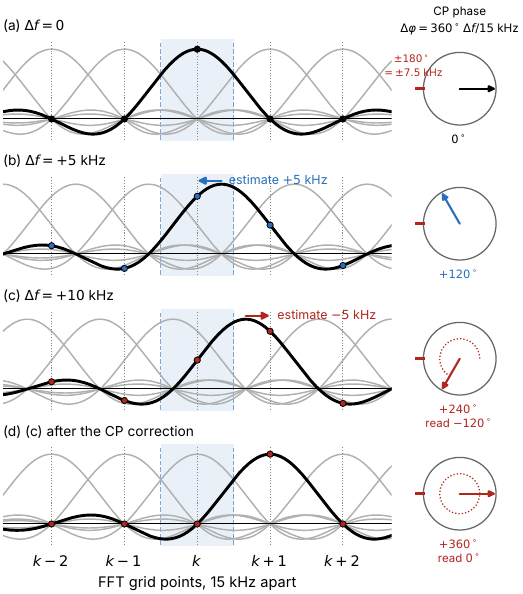}
\caption{Left: subcarrier spectra (sinc) and the FFT sampling points (dots, on subcarrier $k$ in black); shaded: the CP reach of grid point $k$, $\pm$\SI{7.5}{kHz}. Right: the phase difference between the CP and the last samples of the symbol, which the CP copies, $\Delta\varphi=2\pi\Delta f/\Delta f_{\mathrm{sc}}$, read only within $\pm180^\circ$. The CP estimate shifts the spectrum until this phase reads zero, i.e., toward the grid point nearest to the peak (arrow). (a)~$\Delta f=0$: each grid point samples the peak of its own subcarrier and the zeros of the others. (b)~$+\SI{5}{kHz}$: $\Delta\varphi=+120^\circ$ is read correctly; the samples miss the peak (ICI) until the estimate, $+\SI{5}{kHz}$, moves it back to $k$. (c)~$+\SI{10}{kHz}$: $\Delta\varphi=+240^\circ$ is read as $-120^\circ$; the estimate is \SI{-5}{kHz}, and the correction moves the peak to $k+1$. (d)~After this correction, $\Delta\varphi=+360^\circ$ reads as $0^\circ$: the CP sees no offset, but grid point $k+1$ samples the peak of subcarrier $k$.}
\label{fig:wall}
\end{figure}
\end{eoblock}

\subsection{How a Legacy UE Measures Its Neighbors}
LTE handover is network-controlled and UE-assisted \cite{noauthor_ts_nodate-3}. The UE measures the RSRP of its serving cell and of its neighbors: the average power on the resource elements (REs) of the cell-specific reference signals (CRS) \cite{noauthor_ts_nodate-6}. It reports a neighbor through event A3 when
\begin{equation}
M_n - \mathrm{Hys} > M_s + \mathrm{Off}
\label{eq:a3}
\end{equation}
holds for a time-to-trigger (TTT) \cite{noauthor_ts_nodate-5}, where $M_n$ and $M_s$ are the neighbor and serving RSRP in dB, $\mathrm{Off}$ is the A3 offset and $\mathrm{Hys}$ the hysteresis (cell-specific offsets omitted). The offset sets how much stronger the neighbor must be; a negative offset reports a neighbor that is still slightly weaker. The hysteresis and the TTT filter toggling and short fades. A report is thus only as good as $M_n$, and $M_n$ is correct only if the UE is frequency-aligned to the measured cell.

\subsection{Residual Doppler in DtC}
In DtC, every cell reaches the UE with its own residual Doppler~$r$. Ookla's crowdsourced analysis puts the effective beam radius at about \SI{19}{km} at the zenith and \SI{44}{km} at $30^\circ$ elevation \cite{jackson_ookla_2026}; at the beam edge the residual is only 2 to \SI{4}{kHz}, far inside the CP reach. Yet Liu et al.\ measured residuals of up to \SI{10}{kHz} (median \SI{5.3}{kHz}) on the same system in Japan, and the phones still decoded the downlink \cite{liu_variegated_2026}. The UE can hold a cell at such a residual in two ways: its initial cell search includes an integer-subcarrier search, so it can acquire a cell at \SI{10}{kHz}; or it acquires the cell at a small residual, and its automatic frequency control (AFC) tracks the residual as the pass changes it. The carrier the UE settles on for its serving cell is its \emph{serving lock}. Either way, the serving cell is decoded at a large carrier offset, and the question is what this offset does to the neighbor search at the serving lock.

\begin{figure}[t]
\centering
\begin{tikzpicture}[x=0.39cm, y=1cm, font=\scriptsize]
\fill[cT0!12] (-3,0) rectangle (3,0.62);
\draw[-{Latex[length=1.6mm]}, gray!70!black] (-4.6,0) -- (15.7,0) node[above left, inner sep=1pt, text=black]{$f$};
\draw[gray!60, densely dotted] (0,0) -- (0,0.62);
\draw[cT0, line width=1.4pt] (-1,-0.08) -- (-1,0.62);
\draw[cT1, line width=1.4pt] (10,-0.08) -- (10,0.62);
\draw[cT4, densely dashed, thick] (-1,0.3) -- (10,0.3);
\node[cT4, anchor=south, inner sep=1pt] at (6.2,0.3) {$\Delta f=r_n-r_s$};
\node[cT0, anchor=south, inner sep=1pt] at (-0.6,0.62) {neighbor $f_c+r_n$, small $|r_n|$};
\node[cT1, anchor=south, inner sep=1pt] at (10.4,0.62) {serving $f_c+r_s$, $r_s=\SI{10}{kHz}$};
\node[cT4, anchor=west, inner sep=1pt, align=left] at (10.35,0.3) {missed or\\RSRP read low};
\node[anchor=north, inner sep=1.5pt, align=center] at (10,-0.08) {$+10$\\[-1pt]\textcolor{cT1}{serving lock}\\[-1pt]\textcolor{cT4}{neighbor searched here}};
\node[anchor=north, inner sep=1.5pt, align=center] at (0,-0.08) {$f_c$\\[-1pt]\SI{1992.5}{MHz}};
\node[anchor=north east, inner sep=1.5pt, gray!60!black] at (15.7,-0.08) {kHz};
\end{tikzpicture}
\caption{Frequencies at a legacy UE around $f_c=\SI{1992.5}{MHz}$. The UE is locked to its serving cell at the residual $r_s=\SI{10}{kHz}$. A neighbor with a small residual $r_n$ (shaded) is measured at the lock, $\Delta f=r_n-r_s$ away, and is missed or read low.}
\label{fig:mech}
\end{figure}
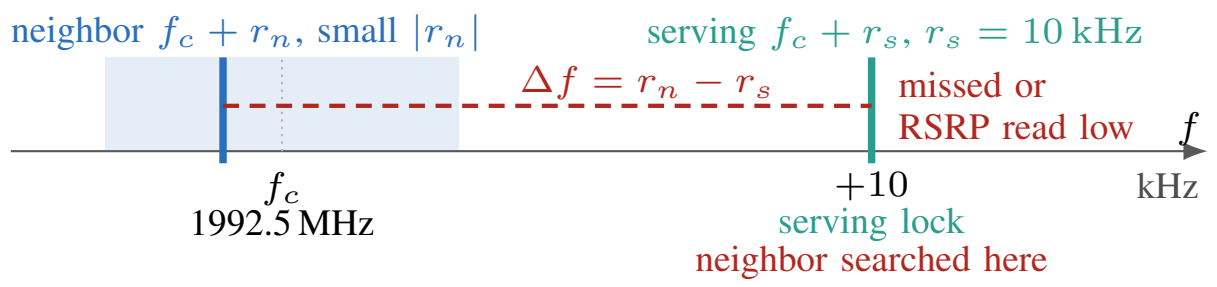

LTE assumes terrestrial cells, which reach the UE on one carrier frequency: an intra-frequency neighbor is a cell on the serving carrier \cite{noauthor_ts_nodate-3}, a measurement object carries one carrier frequency for all its cells \cite{noauthor_ts_nodate-5}, and the measurement requirements set no frequency-offset condition \cite{noauthor_etsi_nodate-1}. A legacy UE keeps its carrier within \SI{0.1}{ppm} of the one received from its serving cell \cite{noauthor_etsi_nodate} and need not search in frequency for a neighbor on that carrier. In DtC, each cell arrives with its own residual, and the assumption fails. If the UE searches and measures its co-channel neighbors at the serving lock, a neighbor with residual $r_n$ arrives at $\Delta f=r_n-r_s$, where $r_s$ is the serving residual (Fig.~\ref{fig:mech}). It loses power to ICI and is read low, and at large offsets it is not identified at all.

Fig.~\ref{fig:geo} shows the two ways a serving beam and a neighbor beam can reach one UE: two beams of one satellite (a), or two beams of two satellites (b). In both, each beam is pre-compensated for its own reference point, $B_s$ for the serving beam and $B_n$ for the neighbor. The UE locks to the frequency of its serving cell, which leaves it the residual $r_s=\fD(\mathrm{UE})-\fD(B_s)$. The neighbor reaches the same UE with its own residual $r_n=\fD(\mathrm{UE})-\fD(B_n)$, so under the serving lock it arrives at the offset $\Delta f=r_n-r_s$.
The terrestrial LTE standard assumes all co-channel cells at the same frequency, and a legacy UE expects its neighbors at its own lock. So, searching neighbor at the serving lock natually gives another frequency offset in DtC.
As a result, its RSRP is read low or it is not found at all, then its report comes late or not at all. The UE stays on the serving cell until the link fails.
The curves below each geometry show where $\Delta f$ comes from.

\begin{eoblock}
\begin{figure*}[t]
\centering
\begin{tikzpicture}[x=0.9cm, y=0.9cm, font=\scriptsize, >={Latex[length=1.6mm]}]
\coordinate (S) at (4.0,2.3);
\coordinate (Bs) at (2.5,0); \coordinate (U) at (3.5,0); \coordinate (Bn) at (4.5,0); \coordinate (G) at (4.0,0);
\draw[gray!70!black, thick] (0.2,0) -- (8.3,0) node[below left=0pt and -2pt, text=gray!60!black, inner sep=1pt] {ground track};
\fill[cT1!25] (S) -- (1.4,0) -- (3.9,0) -- cycle;
\fill[cT0!25] (S) -- (3.1,0) -- (5.6,0) -- cycle;
\fill[cT2!45] (S) -- (3.1,0) -- (3.9,0) -- cycle;
\draw[cT1!70!black, thin] (S) -- (1.4,0) (S) -- (3.9,0);
\draw[cT0!70!black, thin] (S) -- (3.1,0) (S) -- (5.6,0);
\node[text=cT2!70!black, inner sep=1pt, font=\tiny, anchor=south] at (3.5,0.6) {overlap};
\draw[gray!60, densely dotted] (S) -- (G);
\node[draw=gray!60!black, fill=gray!15, rounded corners=1pt, minimum width=5mm, minimum height=2.5mm, inner sep=1pt] at (S) {};
\draw[->, thick] (4.35,2.3) -- (5.3,2.3) node[right, inner sep=2pt] {$\vS$};
\node[above=1.5mm of S, inner sep=1pt] {satellite};
\node[anchor=north west, inner sep=1pt, font=\scriptsize\bfseries] at (0.1,3.0) {(a) two beams of one satellite};
\fill[cT1!70!black] (Bs) circle (1.4pt) node[below=1pt, text=cT1!70!black] {$B_s$};
\fill[cT0!70!black] (Bn) circle (1.4pt) node[below=1pt, text=cT0!70!black] {$B_n$};
\draw[gray!70!black] (G) ++(0,-0.08) -- ++(0,0.16); \node[below=1pt of G, inner sep=1pt, text=gray!60!black] {$G$};
\node[fill=cT4, draw=cT4!70!black, inner sep=1.4pt, label={[text=cT4, inner sep=1pt]above:{UE}}] (ue) at (U) {};
\node[text=cT1!70!black, inner sep=1pt] at (2.0,0.9) {serving};
\node[text=cT0!70!black, inner sep=1pt] at (5.15,0.9) {neighbor};
\draw[->, gray!70!black] (0.2,-2.65) -- (0.2,-0.4) node[above, inner sep=1pt, text=black] {$\fD$};
\draw[gray!60, densely dashed] (0.2,-1.55) -- (8.3,-1.55) node[below left=0pt and -2pt, text=gray!60!black, inner sep=1pt] {$0$};
\draw[thick, gray!50!black, domain=0.4:8.1, samples=60, smooth] plot (\x, {-1.55 + 0.95*tanh((\x-4.0)/1.3)});
\foreach \xx/\col in {2.5/cT1!70!black, 3.5/cT4, 4.5/cT0!70!black} {
  \pgfmathsetmacro{\yy}{-1.55 + 0.95*tanh((\xx-4.0)/1.3)}
  \draw[\col, densely dotted] (\xx,0) -- (\xx,\yy);
  \fill[\col] (\xx,\yy) circle (1.4pt);
  \draw[\col, densely dotted] (\xx,\yy) -- (8.3,\yy);
}
\pgfmathsetmacro{\yBs}{-1.55 + 0.95*tanh((2.5-4.0)/1.3)}
\pgfmathsetmacro{\yU}{-1.55 + 0.95*tanh((3.5-4.0)/1.3)}
\pgfmathsetmacro{\yBn}{-1.55 + 0.95*tanh((4.5-4.0)/1.3)}
\node[right, inner sep=1pt, text=cT1!70!black] at (8.3,\yBs) {$\fD(B_s)$};
\node[right, inner sep=1pt, text=cT4] at (8.3,\yU) {$\fD(\mathrm{UE})$};
\node[right, inner sep=1pt, text=cT0!70!black] at (8.3,\yBn) {$\fD(B_n)$};
\draw[<->, cT4] (7.0,\yBs) -- (7.0,\yU) node[midway, left, inner sep=2pt, text=cT4] {$r_s$};
\draw[<->, cT4] (7.6,\yU) -- (7.6,\yBn) node[midway, left, inner sep=2pt, text=cT4] {$r_n$};
\draw[<->, cT4, thick] (6.3,\yBs) -- (6.3,\yBn) node[midway, left, inner sep=2pt, text=cT4] {$\Delta f$};
\end{tikzpicture}\hfill
\begin{tikzpicture}[x=0.9cm, y=0.9cm, font=\scriptsize, >={Latex[length=1.6mm]}]
\coordinate (S1) at (2.0,2.3); \coordinate (S2) at (5.0,2.3);
\coordinate (Bs) at (2.5,0); \coordinate (U) at (3.5,0); \coordinate (Bn) at (4.5,0); \coordinate (G1) at (2.0,0); \coordinate (G2) at (5.0,0);
\draw[gray!70!black, thick] (0.2,0) -- (8.3,0) node[below left=0pt and -2pt, text=gray!60!black, inner sep=1pt] {ground track};
\fill[cT1!25] (S1) -- (1.4,0) -- (3.9,0) -- cycle;
\fill[cT0!25] (S2) -- (3.1,0) -- (5.6,0) -- cycle;
\fill[cT2!45] (3.1,0) -- (3.9,0) -- (3.5,0.484) -- cycle;
\draw[cT1!70!black, thin] (S1) -- (1.4,0) (S1) -- (3.9,0);
\draw[cT0!70!black, thin] (S2) -- (3.1,0) (S2) -- (5.6,0);
\node[text=cT2!70!black, inner sep=1pt, font=\tiny, anchor=south] at (3.5,0.52) {overlap};
\draw[gray!60, densely dotted] (S1) -- (G1) (S2) -- (G2);
\node[draw=gray!60!black, fill=gray!15, rounded corners=1pt, minimum width=5mm, minimum height=2.5mm, inner sep=1pt] at (S1) {};
\node[draw=gray!60!black, fill=gray!15, rounded corners=1pt, minimum width=5mm, minimum height=2.5mm, inner sep=1pt] at (S2) {};
\draw[->, thick] (2.35,2.3) -- (3.2,2.3) node[right, inner sep=2pt] {$\vS$};
\draw[->, thick] (5.35,2.3) -- (6.2,2.3) node[right, inner sep=2pt] {$\vS'$};
\node[above=1.5mm of S1, inner sep=1pt] {satellite 1};
\node[above=1.5mm of S2, inner sep=1pt] {satellite 2};
\node[anchor=north west, inner sep=1pt, font=\scriptsize\bfseries] at (0.1,3.0) {(b) two beams of two satellites};
\fill[cT1!70!black] (Bs) circle (1.4pt) node[below=1pt, text=cT1!70!black] {$B_s$};
\fill[cT0!70!black] (Bn) circle (1.4pt) node[below=1pt, text=cT0!70!black] {$B_n$};
\draw[gray!70!black] (G1) ++(0,-0.08) -- ++(0,0.16); \node[below=1pt of G1, inner sep=1pt, text=gray!60!black] {$G_1$};
\draw[gray!70!black] (G2) ++(0,-0.08) -- ++(0,0.16); \node[below=1pt of G2, inner sep=1pt, text=gray!60!black] {$G_2$};
\node[fill=cT4, draw=cT4!70!black, inner sep=1.4pt, label={[text=cT4, inner sep=1pt]above:{UE}}] (ue) at (U) {};
\node[text=cT1!70!black, inner sep=1pt] at (1.6,1.1) {serving};
\node[text=cT0!70!black, inner sep=1pt] at (5.6,1.1) {neighbor};
\draw[->, gray!70!black] (0.2,-2.65) -- (0.2,-0.4) node[above, inner sep=1pt, text=black] {$\fD$};
\draw[gray!60, densely dashed] (0.2,-1.55) -- (8.3,-1.55) node[below left=0pt and -2pt, text=gray!60!black, inner sep=1pt] {$0$};
\draw[thick, cT1!60!black, domain=0.4:8.1, samples=60, smooth] plot (\x, {-1.55 + 0.95*tanh((\x-2.0)/1.3)});
\draw[thick, cT0!60!black, domain=0.4:8.1, samples=60, smooth] plot (\x, {-1.55 + 0.95*tanh((\x-5.0)/1.3)});
\node[text=cT1!60!black, inner sep=1pt, anchor=south west] at (0.6,-2.3) {$\fD^{(1)}$};
\node[text=cT0!60!black, inner sep=1pt, anchor=east] at (4.15,-1.95) {$\fD^{(2)}$};
\pgfmathsetmacro{\yBs}{-1.55 + 0.95*tanh((2.5-2.0)/1.3)}
\pgfmathsetmacro{\yUa}{-1.55 + 0.95*tanh((3.5-2.0)/1.3)}
\pgfmathsetmacro{\yBn}{-1.55 + 0.95*tanh((4.5-5.0)/1.3)}
\pgfmathsetmacro{\yUb}{-1.55 + 0.95*tanh((3.5-5.0)/1.3)}
\draw[cT1!70!black, densely dotted] (2.5,0) -- (2.5,\yBs); \fill[cT1!70!black] (2.5,\yBs) circle (1.4pt); \draw[cT1!70!black, densely dotted] (2.5,\yBs) -- (8.3,\yBs);
\draw[cT0!70!black, densely dotted] (4.5,0) -- (4.5,\yBn); \fill[cT0!70!black] (4.5,\yBn) circle (1.4pt); \draw[cT0!70!black, densely dotted] (4.5,\yBn) -- (8.3,\yBn);
\draw[cT4, densely dotted] (3.5,0) -- (3.5,\yUb);
\fill[cT4] (3.5,\yUa) circle (1.4pt); \draw[cT4, densely dotted] (3.5,\yUa) -- (8.3,\yUa);
\fill[cT4] (3.5,\yUb) circle (1.4pt); \draw[cT4, densely dotted] (3.5,\yUb) -- (8.3,\yUb);
\node[right, inner sep=1pt, text=cT4] at (8.3,\yUa) {$\fD^{(1)}(\mathrm{UE})$};
\node[right, inner sep=1pt, text=cT1!70!black] at (8.3,\yBs) {$\fD^{(1)}(B_s)$};
\node[right, inner sep=1pt, text=cT0!70!black] at (8.3,\yBn) {$\fD^{(2)}(B_n)$};
\node[right, inner sep=1pt, text=cT4] at (8.3,\yUb) {$\fD^{(2)}(\mathrm{UE})$};
\draw[<->, cT4] (7.0,\yBs) -- (7.0,\yUa) node[midway, left, inner sep=2pt, text=cT4] {$r_s$};
\draw[<->, cT4] (7.6,\yBn) -- (7.6,\yUb) node[midway, left, inner sep=2pt, text=cT4] {$r_n$};
\end{tikzpicture}
\caption{Doppler of the serving and of the neighbor cell at one UE. (a) Two beams of one satellite: the serving beam is pre-compensated for its reference point $B_s$ and the neighbor beam for $B_n$; the UE sits between the two reference points. Below, the Doppler $\fD$ of a ground point against its position along the track. The UE keeps $r_s=\fD(\mathrm{UE})-\fD(B_s)$ against the serving cell and $r_n=\fD(\mathrm{UE})-\fD(B_n)$ against the neighbor; their difference $\Delta f=r_n-r_s=\fD(B_s)-\fD(B_n)$ is the offset at which the neighbor arrives under the serving lock (Fig.~\ref{fig:mech}), and does not depend on where the UE is. (b) Two beams of two satellites: each cell's Doppler follows its own satellite, $\fD^{(1)}$ for the serving cell and $\fD^{(2)}$ for the neighbor, so the two residuals are read on different curves, $r_s=\fD^{(1)}(\mathrm{UE})-\fD^{(1)}(B_s)$ and $r_n=\fD^{(2)}(\mathrm{UE})-\fD^{(2)}(B_n)$, and their difference $\Delta f=r_n-r_s=\bigl[\fD^{(2)}(\mathrm{UE})-\fD^{(1)}(\mathrm{UE})\bigr]+\bigl[\fD^{(1)}(B_s)-\fD^{(2)}(B_n)\bigr]$ keeps a term in the UE position: the UE sees the two satellites with different Doppler, unlike in (a). Here, $r_s$ and $r_n$ have opposite signs, so $|\Delta f|=|r_s|+|r_n|$.}
\label{fig:geo}
\end{figure*}
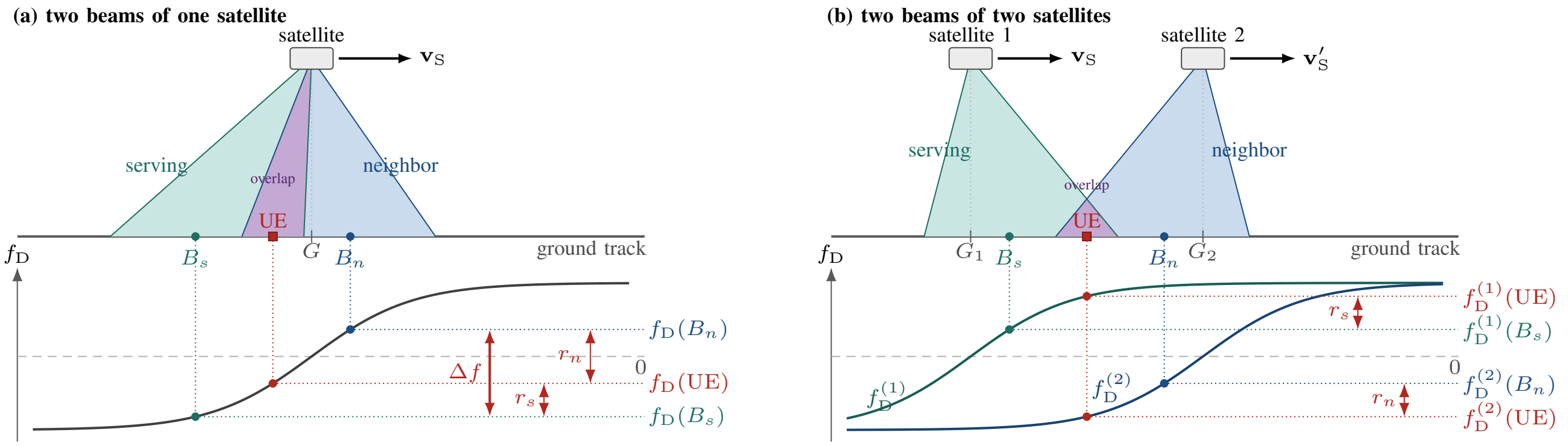
\end{eoblock}

\subsection{Hypothesis}
This paper hypothesizes that \emph{the UE measures its neighbors at the serving lock, so the residual harms not the serving cell, which the UE tracks, but the measurement of its neighbors. In DtC, a serving cell ends on the satellite's schedule, whether the beam is repointed or the satellite moves on; the residual decides only whether the UE has been handed over before then. With the neighbor missed or read low, the handover is missed or late, and the UE stays on the ending cell, with a lower RSRP and a lower MCS, until the link fails.} If this holds, an RLF should come when the serving cell is far below its level while a stronger neighbor exists, and a neighbor should read lower than the same cell read as the serving cell (Section~\ref{sec:field}). In simulation, the neighbor report should fail as the serving residual grows, and UE pre-compensation should restore it (Sections~\ref{sec:sim} and~\ref{sec:results}).

\section{Field Observations: Late Cell Changes}
\label{sec:field}
\subsection{Setup and Data}
All field numbers come from the raw QualiPoc logs of the 28 static DtC sessions with HTTP or iperf traffic of Table~\ref{tab:setup} (345 connected minutes). The phones did not move, so every cell change is set on the satellite side; a stay lasts \SI{23}{s} on average.

\subsection{Most Cell Changes Are Radio Link Failures (RLF)}
From the radio resource control (RRC) messages, which are decoded with pycrate \cite{noauthor_p1secpycrate_2026} against TS~36.331 \cite{noauthor_ts_nodate-5}, every connection receives the same A3 event: a3-Offset \SI{-2.5}{dB}, hysteresis \SI{0.5}{dB} and time-to-trigger \SI{480}{ms}. By~(\ref{eq:a3}), a neighbor is reported once its RSRP exceeds the serving RSRP minus \SI{2}{dB} (the A3 entry) for \SI{480}{ms}.

\begin{figure}[t]
\centering
\includegraphics[width=\columnwidth]{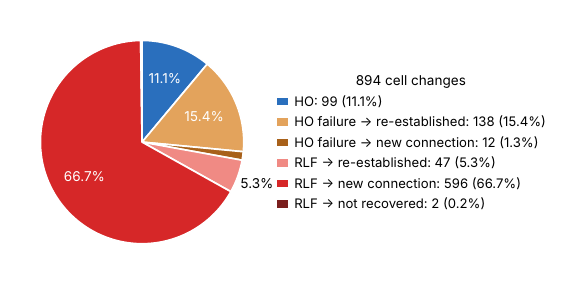}
\caption{Outcome of the 894 cell changes on DtC (345 connected minutes, phones pooled), classified from the RRC messages. HO: the UE received a handover command and completed it. HO failure: the UE received the command but sent a re-establishment request with cause \textit{handoverFailure} within \SI{3}{s} (T304 expired: \SI{100}{ms} from the command, \SI{1000}{ms} on the terrestrial network \cite{noauthor_ts_nodate-5}). RLF: the UE sent a re-establishment request with any other cause, with no handover command before it; mostly T310 expired, \SI{200}{ms} after N310 $=6$ out-of-sync indications \cite{noauthor_ts_nodate-5}. After a failure: re-established, a new connection through idle, or neither (not recovered). The S24+ sends no measurement reports, so all its 329 changes are RLFs.}
\label{fig:changes}
\end{figure}

Fig.~\ref{fig:changes} splits the cell changes by outcome: only \SI{11}{\percent} are handovers (HOs), \SI{17}{\percent} are HO failures, and \SI{72}{\percent} are radio-link failures (RLFs). An HO failure is a commanded handover that the UE did not complete within T304 (Fig.~\ref{fig:changes}), an execution problem beyond this paper. In an RLF, the UE itself declares the link lost, in \SI{88}{\percent} of the RLF reports because the downlink faded until T310 expired (Fig.~\ref{fig:changes}). Of the 645 RLFs, 441 came with no A3 report in the previous \SI{10}{s} and 204 after a report with no command. The S24+ never reports on DtC, though it does on terrestrial LTE, so all its 329 changes are RLFs without a report; of the 316 RLFs of the S25+, 112 came without a report and 204 after one. RLFs are thus the main problem. Under the hypothesis of Section~\ref{sec:background}, both groups share one cause: the residual Doppler makes the UE miss the neighbor or read its RSRP low, so the A3 report comes not at all or late, and the serving link fails before the network can command a handover. The UE then reconnects, by re-establishment on a cell whose eNB holds its context \cite{noauthor_ts_nodate-3} or by a slower new connection.

\subsection{The Serving Cell Degrades Before Each Change}
\begin{figure}[t]
\centering
\includegraphics[width=\columnwidth]{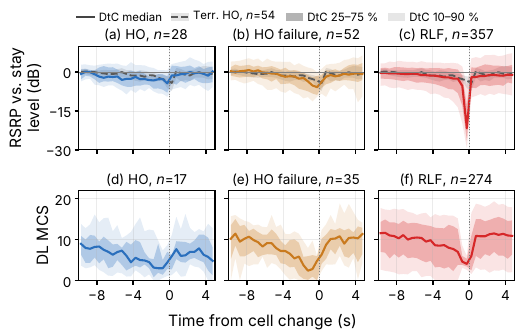}
\caption{Around a cell change on DtC: (a)--(c) serving RSRP vs.\ the stay level of the old cell (dashed: terrestrial handover); (d)--(f) downlink MCS. Old cell before the change, new cell after it; stays $>$\SI{10}{s}.}
\label{fig:drop}
\end{figure}

A cell change comes either from a handover or from a radio link failure (RLF). On the terrestrial network, every cell change was a handover; on DtC, cell changes were a mix of both. Fig.~\ref{fig:drop} follows the serving RSRP and the downlink MCS through each cell change; the stay level is the median RSRP of a stay without its last \SI{10}{s}. RLFs bring the deepest RSRP drop, and they are most of the cell changes. Before an RLF, the serving RSRP stays near its stay level until about \SI{1}{s} before the change, and then collapses (Fig.~\ref{fig:drop}c). In the last second before an RLF, the serving RSRP is \SI{13.2}{dB} below its stay level, at \SI{-135}{dBm} in the median, and the downlink MCS falls from about 11 to 4 (Fig.~\ref{fig:drop}f). Before a successful or failed handover, it is only \SI{3.1}{dB} or \SI{5.2}{dB} below, close to the \SI{4.3}{dB} before a terrestrial handover. In every case the new cell starts at a level comparable to the old stay level: a good cell was available. A handover, when it happens, comes in time, as on the terrestrial network; an RLF moves the UE to that cell only after the serving link has failed.

\subsection{The New Cell Was a Known Neighbor, Read Low Until It Was Too Late}
\begin{figure}[t]
\centering
\includegraphics[width=\columnwidth]{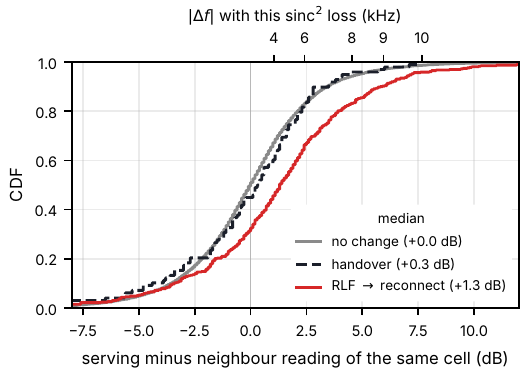}
\caption{RSRP of the same cell read twice, about \SI{0.5}{s} apart: first as a neighbor, then as the serving cell. After a successful handover ($n=98$), the two readings agree (median $+\SI{0.3}{dB}$): the neighbor was read correctly, so its report could trigger the handover in time. After an RLF and a reconnection to that cell ($n=458$), the serving reading is higher (median $+\SI{1.3}{dB}$, P90 $+\SI{5.8}{dB}$): as a neighbor, the cell was read low. No change ($n=13\,751$): the same cell read twice as a neighbor at the same gap, more than \SI{5}{s} from any cell change (median \SI{0.0}{dB}). Top axis: the frequency offset $|\Delta f|$ whose ICI loss alone would cause this gap.}
\label{fig:gap}
\end{figure}

\begin{figure}[t]
\centering
\includegraphics[width=\columnwidth]{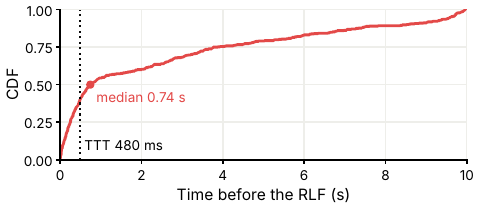}
\caption{Time before the RLF at which the target first rose above the A3 entry (target minus serving RSRP $>\SI{-2}{dB}$) in the last \SI{10}{s}, for the 463 RLFs whose target was measured as a neighbor (phones pooled). Dotted: time-to-trigger. The RLF time is the re-establishment request; the UE declares the failure when T310 (\SI{200}{ms}, started after N310 $=6$ out-of-sync indications) expires, so the link had been out of sync for at least \SI{0.26}{s} before it.}
\label{fig:late}
\end{figure}

In \SI{71.8}{\percent} of all RLFs, the UE had measured the new cell as a neighbor before the failure; \SI{19.2}{\percent} ended on a cell it had never measured. Fig.~\ref{fig:visible}a splits all 645 RLFs by the cell the UE reconnects to: in 412 (\SI{63.9}{\percent}) the new cell was the strongest neighbor the UE had measured in the last \SI{1.5}{s}, in 51 (\SI{7.9}{\percent}) a measured neighbor but not the strongest, and 58 (\SI{9.0}{\percent}) ended on the same cell or on none. At its last neighbor reading, the target led the serving cell by \SI{17.8}{dB} in the median, and only 7 of the 458 targets with a later serving reading were below the A3 entry of \SI{-2}{dB} (Fig.~\ref{fig:visible}b); this lead, however, is the lead of the last second, in which the serving cell falls \SI{13}{dB} (Fig.~\ref{fig:drop}c). As a neighbor, that cell was read low: once it became the serving cell, it read \SI{1.3}{dB} higher in the median and \SI{5.8}{dB} higher at the P90, against \SI{0.0}{dB} for the same cell read twice as a neighbor (Fig.~\ref{fig:gap}). Read low, it rose above the A3 entry only \SI{0.74}{s} (median) before the RLF (Fig.~\ref{fig:late}), too late for the time-to-trigger, the report and the handover command. In \SI{39}{\percent} of these RLFs it rose above the entry less than the \SI{480}{ms} time-to-trigger before the failure, and in only \SI{21}{\percent} more than \SI{5}{s} before; even \SI{0.74}{s} is too little for the layer-3 filter, the time-to-trigger, and the report and the handover command over the satellite link.

\begin{eoblock}
\begin{figure*}[t]
\centering
\includegraphics[width=\textwidth]{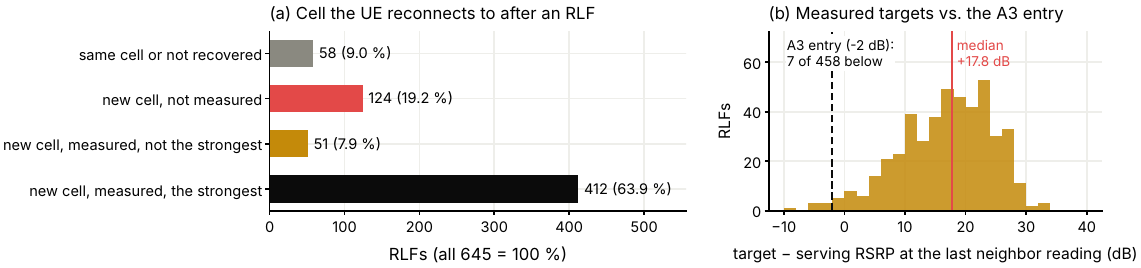}
\caption{Where the UE reconnects after an RLF (phones pooled). (a) All 645 RLFs, split by the cell the UE reconnects to; the four shares add up to \SI{100}{\percent}. Measured: seen as a neighbor in the last \SI{1.5}{s}; strongest: the strongest of these neighbors. (b) Target minus serving RSRP at the target's last neighbor reading, for the 458 targets with a serving reading after the reconnection; dashed: A3 entry.}
\label{fig:visible}
\end{figure*}
\end{eoblock}

\begin{table}[t]
\centering
\caption{Field Measurement Setup}
\label{tab:setup}
\begin{tabular}{@{}ll@{}}
\toprule
Phones & Galaxy S24+, S25+ forced to EARFCN 8665 (\SI{1992.5}{MHz})\\
Traffic & SigCap \cite{dogan-tusha_spectrum_2025}: iperf/HTTP; neighbor PCIs every \SI{5}{s}\\
PHY log & QualiPoc \cite{rohde__schwarz_qualipoc_nodate} raw log; SmartAnalytics \cite{rohde__schwarz_gmbh__co_kg_smartanalytics_nodate} CSV\\
Sites & Static sites on Notre Dame campus, Jul.--Oct.\ 2026\\
Terrestrial & Same phone models, T-Mobile NR n25\\
\bottomrule
\end{tabular}
\end{table}

\section{Simulation Setup}
\label{sec:sim}
\begin{figure}[t]
\centering
\includegraphics[width=0.85\columnwidth]{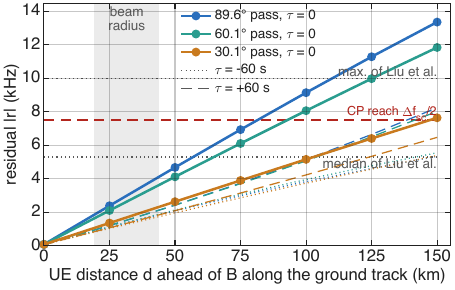}
\caption{Simulation~1: residual $|r|$ against the UE distance $d$ ahead of $B$ on the three passes, at the closest approach ($\tau=0$) and at $\tau=\pm\SI{60}{s}$. Shaded: the effective beam radius of \cite{jackson_ookla_2026}. Dashed: the CP reach; dotted: the median and the maximum residual measured by \cite{liu_variegated_2026}.}
\label{fig:rvsd}
\end{figure}

The logs of Section~\ref{sec:field} hold no frequency offset, so the residual behind the late cell changes is not observed. Simulation~1 supplies it for the orbit of the system and the site of the measurements: the residual $r$ of a UE at a distance $d$ from the reference point of its cell (Fig.~\ref{fig:rvsd}), and what $r$ does to a cell decoded with a CP estimate, in an LTE downlink chain. Simulation~2 brings both to a serving cell and a neighbor measured at the serving lock.

\begin{table}[t]
\centering
\caption{Simulation~1: Orbit, Geometry and Snapshots}
\label{tab:geo}
\begin{tabular}{@{}ll@{}}
\toprule
Orbit & \SI{525}{km}, $53^\circ$ \cite{space_exploration_technologies_corp_spacex_sched_2025}\\
Carrier, max.\ Doppler & \SI{1992.5}{MHz}, $f_{\max}=\SI{48.5}{kHz}$\\
Reference point $B$ & Notre Dame, $41.7034^\circ$\,N, $86.2390^\circ$\,W\\
Peak elevation of 3 Passes & $89.6^\circ$, $60.1^\circ$ (asc.), $30.1^\circ$ (desc.)\\
7 UE distances $d$ ahead of $B$ & $0,25,\dots,\SI{150}{km}$\\
3 Instants $\tau$ from closest appr. & $0,\pm\SI{60}{s}$\\
Geometry points & $3\times7\times3=63$\\
\bottomrule
\end{tabular}
\end{table}

\begin{figure}[t]
\centering
\IfFileExists{figs/paper_fig_pass.pdf}{\includegraphics[width=\columnwidth]{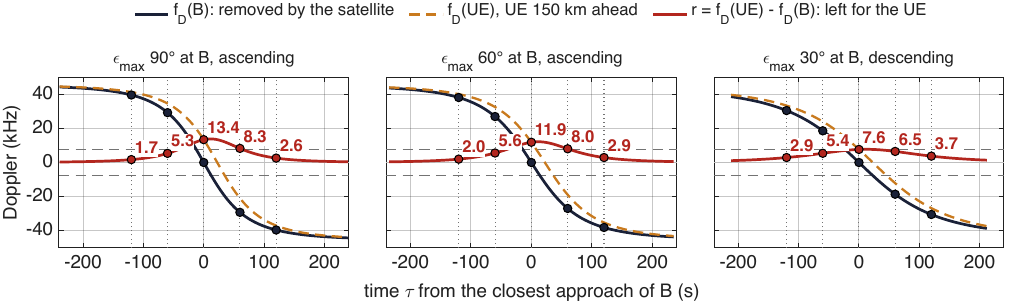}}{\includegraphics[width=\columnwidth]{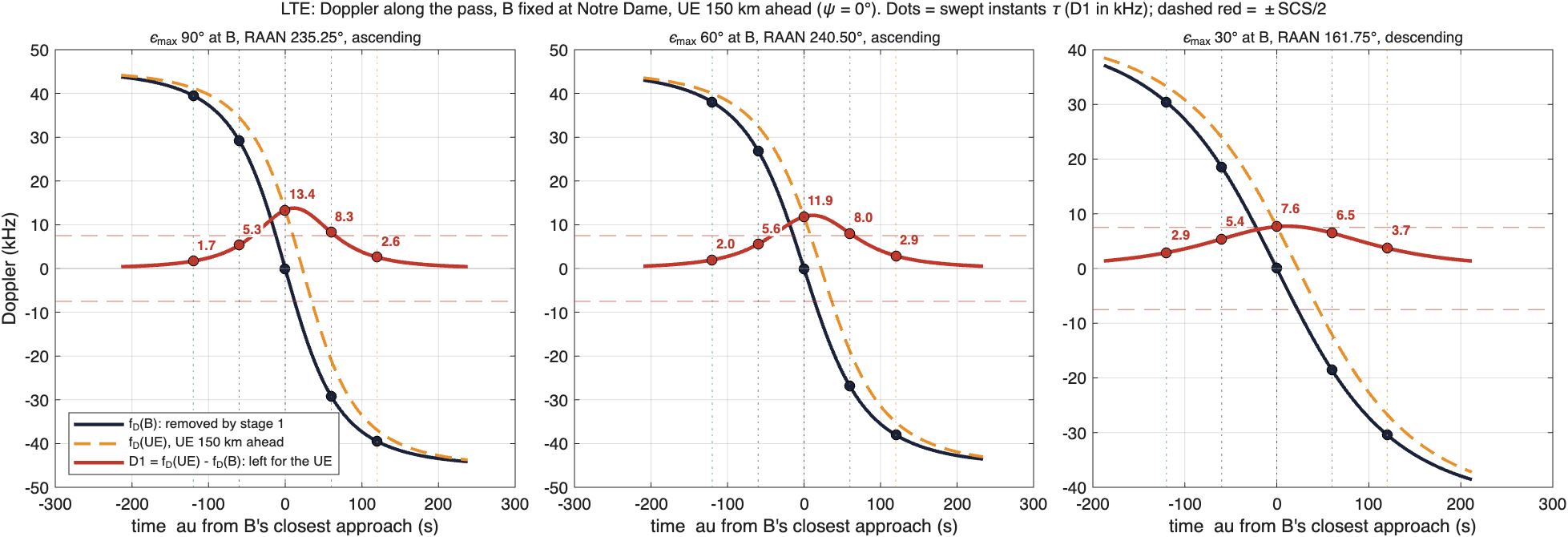}}
\caption{Doppler along the three passes of Table~\ref{tab:geo}, against the time $\tau$ from the closest approach of $B$, for a UE \SI{150}{km} ahead of $B$ along the ground track: the Doppler $\fD(B)$ that the satellite pre-compensates (solid black), the Doppler $\fD(\mathrm{UE})$ at the UE (dashed orange) and the residual $r=\fD(\mathrm{UE})-\fD(B)$ left at the UE (solid red). Dots: the swept instants, labeled with $r$ in kHz. Dashed gray: the CP reach $\pm\Delta f_{\mathrm{sc}}/2$.}
\label{fig:pass}
\end{figure}

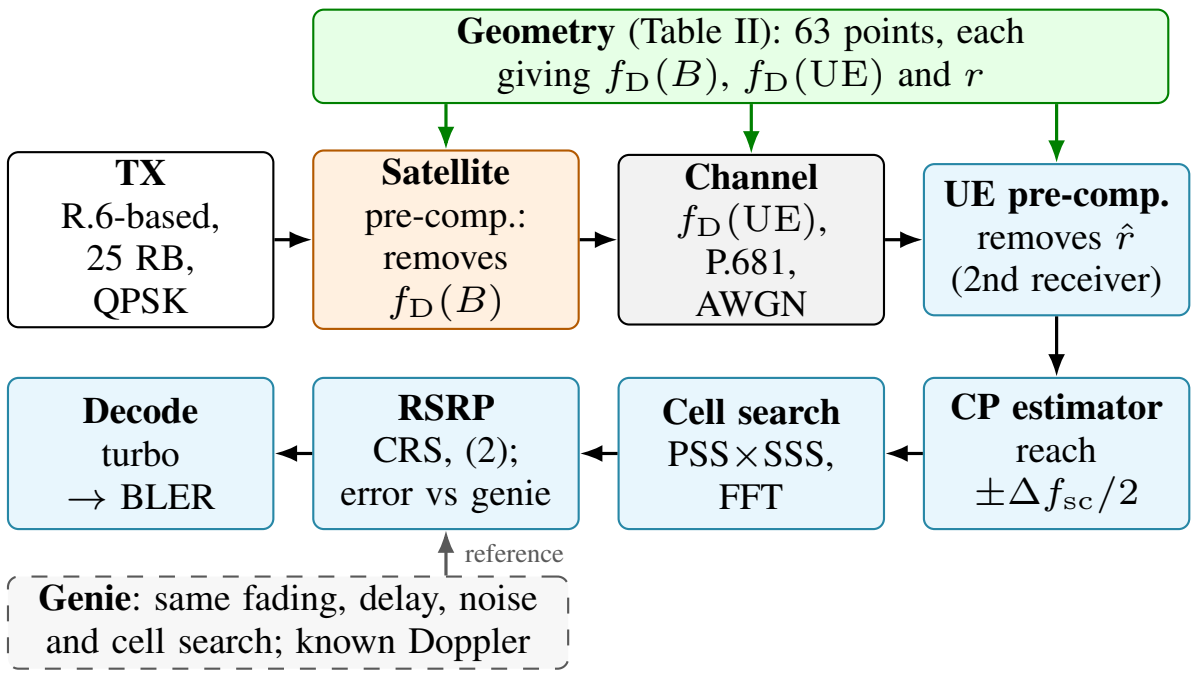
\begin{figure}[t]
\centering
\begin{tikzpicture}[
  node distance=2.5mm and 2.5mm, font=\scriptsize,
  blk/.style={draw, rounded corners=2pt, align=center, minimum height=10mm, text width=16.5mm, inner sep=1.5pt},
  sat/.style={blk, fill=orange!12, draw=orange!70!black},
  ue/.style={blk, fill=cyan!8, draw=cyan!60!black},
  ch/.style={blk, fill=gray!10},
  geo/.style={blk, fill=green!10, draw=green!50!black, minimum height=0pt},
  arr/.style={-{Latex[length=1.5mm]}, semithick},
  garr/.style={-{Latex[length=1.5mm]}, semithick, green!50!black}]
\node[blk] (tx) {\textbf{TX}\\ R.6-based, 25 RB,\\ QPSK};
\node[sat, right=of tx] (s1) {\textbf{Satellite}\\ pre-comp.:\\ removes $\fD(B)$};
\node[ch, right=of s1] (chn) {\textbf{Channel}\\ $\fD(\mathrm{UE})$, P.681,\\ AWGN};
\node[ue, right=of chn] (s2) {\textbf{UE pre-comp.}\\ removes $\hat r$\\ (2nd receiver)};
\node[ue, below=4mm of s2] (cp) {\textbf{CP estimator}\\ reach\\ $\pm\Delta f_{\mathrm{sc}}/2$};
\node[ue, left=of cp] (sync) {\textbf{Cell search}\\ PSS$\times$SSS,\\ FFT};
\node[ue, left=of sync] (rsrp) {\textbf{RSRP}\\ CRS, \eqref{eq:rsrp};\\ error vs genie};
\node[ue, left=of rsrp] (dec) {\textbf{Decode}\\ turbo\\ $\rightarrow$ BLER};
\node[geo, above=3mm of s1.north west, anchor=south west, text width=55.5mm] (geo) {\textbf{Geometry} (Table~\ref{tab:geo}): 63 points, each giving $\fD(B)$, $\fD(\mathrm{UE})$ and $r$};
\draw[arr] (tx)--(s1); \draw[arr] (s1)--(chn); \draw[arr] (chn)--(s2); \draw[arr] (s2)--(cp);
\draw[arr] (cp)--(sync); \draw[arr] (sync)--(rsrp); \draw[arr] (rsrp)--(dec);
\draw[garr] (geo.south -| s1) -- (s1.north);
\draw[garr] (geo.south -| chn) -- (chn.north);
\draw[garr] (geo.south -| s2) -- (s2.north);
\node[blk, dashed, fill=gray!6, draw=gray!70!black, below=3mm of dec.south west, anchor=north west, text width=36mm, minimum height=0pt] (gen) {\textbf{Genie}: same fading, delay, noise and cell search; known Doppler};
\draw[arr, gray!70!black] (gen.north -| rsrp) -- (rsrp.south) node[midway, right, font=\tiny] {reference};
\end{tikzpicture}
\caption{Simulation~1 workflow. Green: geometry; orange: satellite; blue: UE. Gray: genie, the RSRP reference. The legacy receiver skips UE pre-compensation and meets the full $r$ at the CP estimator; the second receiver removes $\hat r=r$ first.}
\label{fig:pipe}
\end{figure}

\begin{figure}[t]
\centering
\begin{tikzpicture}[
  node distance=2.5mm and 2.5mm, font=\scriptsize,
  blk/.style={draw, rounded corners=2pt, align=center, minimum height=10mm, text width=16.5mm, inner sep=1.5pt},
  ue/.style={blk, fill=cyan!8, draw=cyan!60!black},
  ch/.style={blk, fill=gray!10},
  arr/.style={-{Latex[length=1.5mm]}, semithick}]
\node[blk] (stx) {\textbf{Serving cell}\\ PCI~0};
\node[ch, right=of stx] (sch) {\textbf{Channel}\\ residual $r_s$,\\ P.681, AWGN};
\node[ue, right=of sch] (lock) {\textbf{Serving lock}\\ $f_{\mathrm{lock}}$: $r_s$ plus\\ CP estimate};
\node[ue, right=of lock] (srs) {\textbf{Serving}\\ search, RSRP};
\node[blk, below=5mm of stx] (ntx) {\textbf{Neighbor cell}\\ PCI~1, $+$\SI{3}{dB}};
\node[ch, right=of ntx] (nch) {\textbf{Channel}\\ residual $r_n$, delay,\\ P.681, AWGN};
\node[ue, right=of nch] (der) {\textbf{De-rotate}\\ legacy: $f_{\mathrm{lock}}$\\ 2nd: $r_n$, CP};
\node[ue, right=of der] (nrs) {\textbf{Neighbor}\\ PSS$\times$SSS,\\ RSRP};
\node[ue, below=3mm of nrs] (a3) {\textbf{A3 report}\\ neighbor $-$\\ serving $>\SI{-2}{dB}$};
\node[blk, dashed, fill=gray!6, draw=gray!70!black, left=of a3, text width=36mm, minimum height=0pt] (gen) {\textbf{Genie} for both cells: same fading, delay, noise and search; known Doppler};
\draw[arr] (stx)--(sch); \draw[arr] (sch)--(lock); \draw[arr] (lock)--(srs);
\draw[arr] (ntx)--(nch); \draw[arr] (nch)--(der); \draw[arr] (der)--(nrs);
\draw[arr] (lock)--(der) node[midway, right, font=\tiny] {$f_{\mathrm{lock}}$};
\draw[arr] (nrs)--(a3);
\draw[arr] (srs.east) -- ++(1.5mm,0) |- (a3.east);
\draw[arr, gray!70!black] (gen)--(a3);
\end{tikzpicture}
\caption{Simulation~2 workflow. Gray: channels; blue: UE. The legacy receiver de-rotates the neighbor by the serving lock, so it arrives at $\Delta f=r_n-r_s$; the UE pre-compensation receiver removes $r_n$ first. Dashed: genie, the reference for the RSRP error and the report.}
\label{fig:pipe2}
\end{figure}
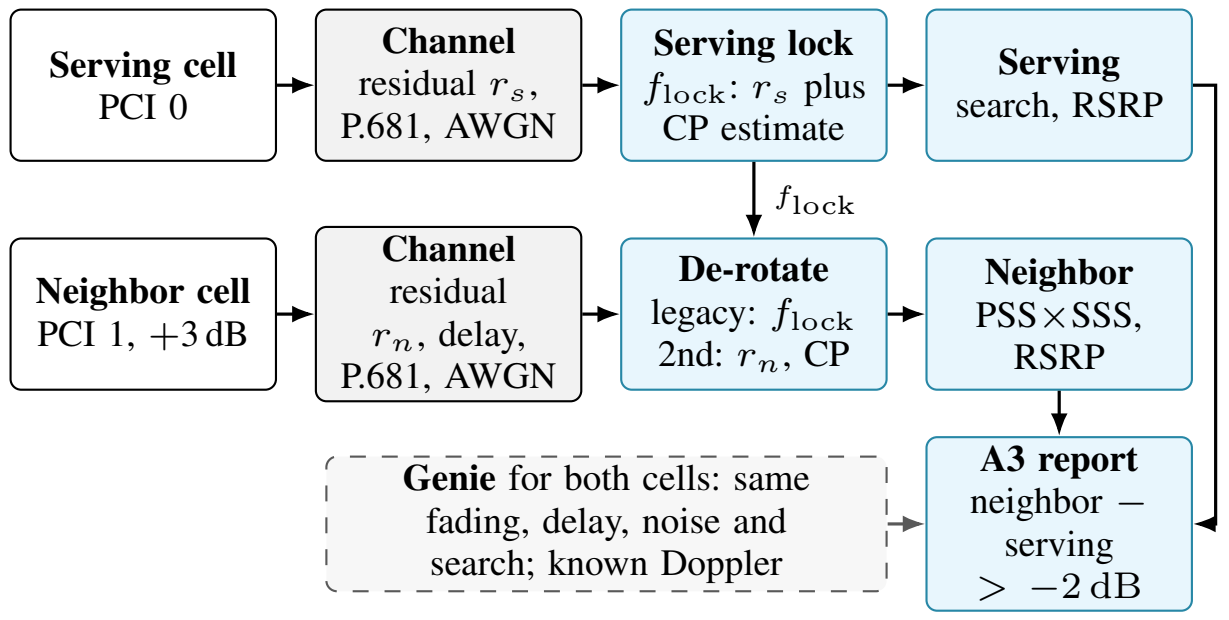

\subsection{Orbit and Residual Doppler}
Table~\ref{tab:geo} lists the orbit, the geometry and the snapshots, and Fig.~\ref{fig:pipe} shows the workflow. The satellite pre-compensates each cell for the Doppler shift $\fD(B)$ at its reference point $B$, the beam center. A UE elsewhere in the cell keeps the residual $r=\fD(\mathrm{UE})-\fD(B)$, where $\fD(X)=-(f_c/c)\,\vS\cdot\mathbf{u}_X$ for a ground point $X$ seen along the unit vector $\mathbf{u}_X$. The UE is placed ahead of $B$ along the ground track, where $r$ is largest; Fig.~\ref{fig:rvsd} shows $r$ against $d$. Inside the effective beam radius of \cite{jackson_ookla_2026}, $r$ stays below about \SI{4}{kHz} on every pass. Near the closest approach, $r$ reaches the CP reach of \SI{7.5}{kHz} at $d=82$, 94 and \SI{149}{km} on the three passes, and a residual of \SI{10}{kHz}, the largest that \cite{liu_variegated_2026} measured, corresponds to about \SI{110}{km} on the overhead pass. Inside its beam, the serving cell is thus not the problem; a UE reaches the residuals measured in the field on a cell whose reference point is far away, whether its AFC has followed the cell there or that cell was the best link, and it measures each neighbor at this lock.

Fig.~\ref{fig:pass} follows the Doppler along each pass for the farthest UE of Table~\ref{tab:geo}. At the closest approach of the beam center $B$, the velocity of the satellite is perpendicular to the line from the satellite to $B$, so $\fD(B)$, and with it the pre-compensation, is zero; the UE keeps its whole Doppler as the residual. On a high pass the satellite is then almost overhead, where the Doppler changes fastest across the beam, so this residual is largest: \SI{13.4}{kHz} on the overhead pass, \SI{11.9}{kHz} on the $60^\circ$ pass and \SI{7.6}{kHz} on the $30^\circ$ pass.

\subsection{Single-Cell Link-Level Chain}
\label{sec:pipe}
Each geometry point drives an LTE downlink chain built with the MATLAB LTE Toolbox (Fig.~\ref{fig:pipe}).

\textit{Transmitter.} The reference channel is based on R.6, with 25 resource blocks (\SI{5}{MHz}), one CRS port and normal CP. QPSK runs at per-RE signal-to-noise ratios (SNRs) of $-3$, 0, 3, 6 and \SI{9}{dB}, with the code rates 0.08, 0.12, 0.19, 0.48 and 0.59 of the CQI table. These values follow the field data of Section~\ref{sec:field}.

\textit{Channel.} The satellite pre-compensates $\fD(B)$. The channel then applies the true $\fD(\mathrm{UE})$, ITU-R P.681 land-mobile-satellite fading \cite{noauthor_recommendation_nodate} and AWGN. The fading uses the Suburban environment at the true UE elevation, drawn afresh per subframe.

\textit{Receivers.} Two receivers process every point. Both receivers estimate the frequency offset with the blind CP correlator \cite{van_de_beek_ml_1997} over one subframe, and de-rotate; then they search the cell with its primary and secondary synchronization signals (PSS, SSS), which also give the frame timing.  The \emph{UE pre-compensation} receiver first removes $\hat r=r$, computed from exact ephemeris of the satellite, its exact position and $B$. It is an upper bound on any UE-side correction.

\textit{RSRP.} On each CRS port-0 RE, the UE multiplies the received value by the conjugate of the known CRS symbol. With $L=2$ CRS symbols on each of the $K$ CRS subcarriers and the noise estimate $\hat\sigma^2$,
\begin{equation}
\mathrm{RSRP}=\frac{1}{K}\sum_{k}\biggl(\Bigl|\frac{1}{L}\sum_{l}Y_{k,l}S_{k,l}^{*}\Bigr|^{2}-\frac{\hat\sigma^{2}}{L}\biggr).
\label{eq:rsrp}
\end{equation}
On the right REs this returns the channel power. On the wrong REs, each term carries a data symbol with a random phase, the noise estimate counts the whole signal, and the RSRP falls to the noise floor. The RSRP error is the measured RSRP minus that of a genie with the same fading, delay and noise but a known Doppler, which runs the same cell search and \eqref{eq:rsrp} with its own noise estimate. It is averaged over 25 subframes.

\textit{Decoding.} The data channel is turbo-decoded without retransmissions; a loss of synchronization counts as a block error in the block error rate (BLER).

\subsection{Two-Cell Setup: The Neighbor Under the Serving Lock}
\label{sec:twocell}
Simulation~2 brings the chain and the residuals of Simulation~1 to a serving cell and a neighbor: it has one UE and two cells. The orbits are left out, so the two cells may be two beams of one satellite or of two satellites; instead of a geometry, each cell's residual to its own center is set directly. The serving cell (PCI~0) arrives with the residual $r_s$ and the neighbor (PCI~1) with the residual $r_n$ to its own center. Both carry the \SI{5}{MHz} QPSK downlink of Section~\ref{sec:pipe}. Fig.~\ref{fig:pipe2} shows the workflow.

\textit{Serving lock.} A converged tracking loop is modeled: it removes $r_s$ and adds one CP estimate on the serving subframe. The lock $f_{\mathrm{lock}}$ thus differs from $r_s$ only by the CP estimator's noise. The serving cell is searched and measured there.

\textit{Neighbor.} Under the serving lock, the received neighbor is de-rotated by $f_{\mathrm{lock}}$. It therefore arrives at the offset
\begin{equation}
\Delta f=r_n-r_s
\label{eq:offset}
\end{equation}
plus the lock error. The neighbor's own residual is $r_n$. The serving residual $r_s$ is swept from 0 to \SI{15}{kHz} in \SI{1}{kHz} steps. The UE correlates a \SI{3}{ms} stream with the neighbor's PSS and SSS, with the timing unknown within $\pm$\SI{100}{\micro\second}. The neighbor is identified when the PSS$\times$SSS metric exceeds 4.5. Its RSRP is the mean CRS power per RE in one subframe, minus a noise estimate from the differences of adjacent CRS subcarriers.

\textit{Receivers.} Two receivers measure the neighbor on the same samples. The \emph{legacy} receiver searches and measures it under the serving lock, without a frequency estimate of its own. The \emph{UE pre-compensation} receiver removes the true $r_n$ before the CP estimator and the search. It is an upper bound.

\textit{Report.} The neighbor is \emph{reported} when its measured RSRP exceeds the serving RSRP by more than the a3-Offset plus the hysteresis \cite{noauthor_ts_nodate-5}, \SI{-2}{dB} on the US network (Section~\ref{sec:field}). Each trial is one subframe, so the time-to-trigger and the layer-3 filter are not modeled. The genie report uses the genie of Section~\ref{sec:pipe} for both cells.

\textit{Conditions.} The serving per-RE SNR is $-3$, 0, 3, 6 or \SI{9}{dB}, and the neighbor is \SI{3}{dB} stronger. Each cell has its own P.681 Suburban fading at an elevation of $90^\circ$, $60^\circ$ or $30^\circ$, drawn afresh in each of 40 trials per point.

\begin{figure}[t]
\centering
\includegraphics[width=\columnwidth]{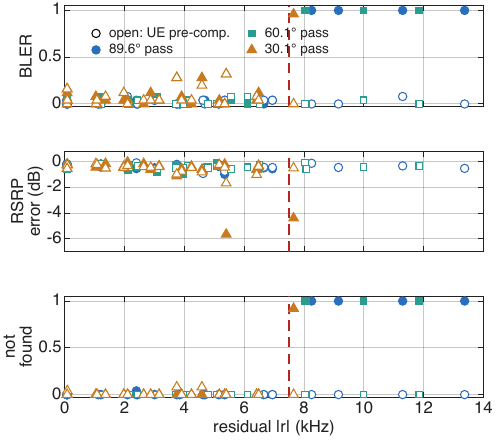}
\caption{Simulation~1 at \SI{3}{dB} SNR: BLER (top), RSRP error (middle) and the fraction of subframes in which the cell is not found (bottom), for the legacy receiver (filled) and the UE pre-compensation receiver (open), against the residual $|r|$, one marker per geometry point of Table~\ref{tab:geo}. Dashed: the CP reach. A legacy point whose cell is not found has no RSRP and is missing from the middle panel.}
\label{fig:sim1}
\end{figure}

\begin{figure*}[!tb]
\centering
\includegraphics[width=0.85\textwidth]{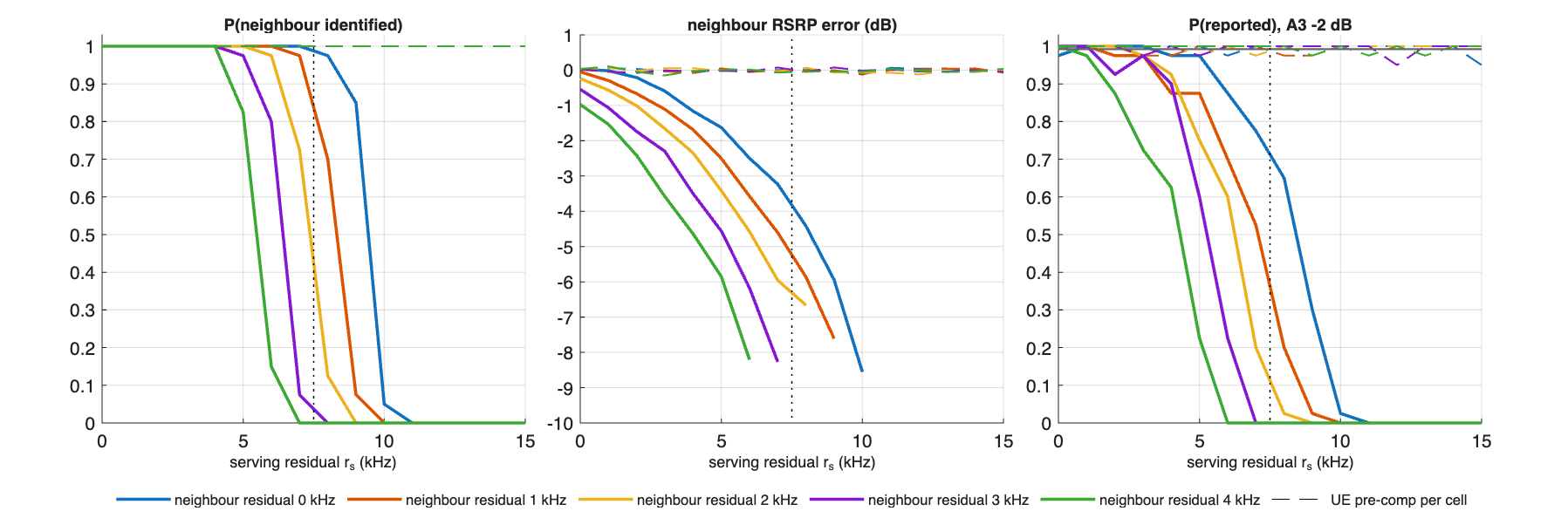}
\caption{Simulation~2, serving SNR \SI{3}{dB}, neighbor \SI{3}{dB} stronger: identification, RSRP error and report (A3 entry \SI{-2}{dB}) of the neighbor against the serving residual $r_s$, for a neighbor residual $|r_n|$ of 0 to \SI{4}{kHz} with the UE between the two reference points. Solid: legacy receiver; dashed: UE pre-compensation per cell. Dotted vertical: $\Delta f_{\mathrm{sc}}/2$. Gray horizontal: genie report rate.}
\label{fig:twocell}
\end{figure*}

\section{Results}
\label{sec:results}
\subsection{Single Cell: The CP Reach}
\label{sec:res1}
The hypothesis of Section~\ref{sec:background} puts the harm not on the serving cell but on the neighbor; Simulation~1 checks the first half. Fig.~\ref{fig:sim1} shows the BLER, the RSRP error and the share of subframes in which the cell is not found, against the residual $|r|$. Below $\pm$\SI{7.5}{kHz} the two receivers coincide: at \SI{3}{dB} SNR, the legacy receiver has a mean BLER of 0.05 and a median RSRP error of \SI{-0.36}{dB}, against 0.04 and \SI{-0.42}{dB} with UE pre-compensation; the same holds at every SNR. A residual inside the reach does not harm a cell, and the serving cell, which the UE tracks, stays inside it: in the field, the serving RSRP holds its stay level until the cell ends (Section~\ref{sec:field}).

Above $\pm$\SI{7.5}{kHz} the legacy receiver loses the cell at all nine points. At \SI{3}{dB} SNR it does not find the cell in \SI{99}{\percent} of the subframes, and the BLER is at least 0.96. Where it is still found, mostly at 6 and \SI{9}{dB} SNR, the CRS is read from the wrong REs, and \eqref{eq:rsrp} reads 4 to \SI{11}{dB} low. With UE pre-compensation, the nine points return to a BLER of 0.01 and an RSRP error of \SI{-0.39}{dB}.

\subsection{Two Cells: Identification, RSRP and Report}
\label{sec:res2}
Simulation~2 checks the second half: that the serving lock hides the neighbor. Fig.~\ref{fig:twocell} shows the neighbor against the serving residual $r_s$. The lock itself is sound: the median lock error is below \SI{410}{Hz} at \SI{-3}{dB} SNR and below \SI{50}{Hz} at \SI{9}{dB}, and the serving RSRP error shows no trend with $r_s$.

\textit{The same lock moves the neighbor off the grid.} With $r_n=0$, the neighbor arrives at $\Delta f=-r_s$. The neighbor is read low since the noise estimate subtracts the ICI (Section~\ref{sec:background}), so the reading is low by $10\log_{10}\mathrm{sinc}^2(\Delta f/\Delta f_{\mathrm{sc}})$: \SI{-1.7}{dB} at \SI{5}{kHz} and \SI{-4.5}{dB} at \SI{8}{kHz}, within \SI{0.3}{dB} of theory. This is the curve on which the field reading gap of Fig.~\ref{fig:gap} was read: \SI{1.3}{dB} at \SI{4.5}{kHz}, \SI{5.8}{dB} at \SI{8.9}{kHz}.

\textit{Under the \SI{-2}{dB} entry, the report fails before the identification.} The genie reports the neighbor in \SI{99}{\percent} of the trials. The entry leaves \SI{5}{dB} of headroom, which sinc$^2$ uses up only near \SI{8.4}{kHz}. At \SI{3}{dB} SNR the legacy receiver reports the neighbor in 97, 88, 78, 65, 30 and \SI{3}{\percent} of the trials at 5, 6, 7, 8, 9 and \SI{10}{kHz}. Up to \SI{9}{kHz} the neighbor is found but read below the entry: the UE knows it, yet does not report it until the serving cell collapses and lifts it over the entry, the \SI{0.74}{s} of Section~\ref{sec:field}. Beyond, it is almost never found, like the new cell that the UE had never measured.

\textit{A neighbor with its own residual fails earlier.} With the UE between the two reference points, $r_s$ and $r_n$ have opposite signs, so $|\Delta f|=r_s+|r_n|$ and every curve shifts left by $|r_n|$ (Fig.~\ref{fig:twocell}). At $r_s=\SI{5}{kHz}$, the median of \cite{liu_variegated_2026}, the neighbor is read 1.6, 2.5, 3.4, 4.6 and \SI{5.9}{dB} low for $|r_n|=0$ to \SI{4}{kHz}, and its report falls from 97 to 88, 75, 60 and \SI{23}{\percent}. At $r_s=\SI{10}{kHz}$, the maximum of \cite{liu_variegated_2026}, the neighbor is found in only \SI{5}{\percent} of the trials at $r_n=0$ and in none for $|r_n|\geq\SI{1}{kHz}$, so it is almost never reported. Without a report, the network cannot command a handover, and the UE stays on its serving cell until the link fails: the RLFs of Section~\ref{sec:field}, 441 of them with no report at all.

\textit{UE pre-compensation reads the neighbor correctly.} At \SI{3}{dB} SNR, for every $r_s$ and $r_n$, it reads the neighbor within \SI{0.15}{dB} of the genie and reports it as often, within \SI{3}{\percent} points. Both predictions of Section~\ref{sec:background} hold: the neighbor report fails as the serving residual grows, and pre-compensation per cell restores it.

\section{Discussion}
\label{sec:discussion}
\subsection{Delivering the Per-Cell Correction to a Legacy Phone}
How the reference point of each cell reaches a legacy phone is open. In NR Release~17, SIB19 gives NTN-capable UEs the ephemeris of the serving and neighbor satellites \cite{noauthor_ts_nodate}, but not the reference point for which the network pre-compensates each cell. 

The alternative needs no signaling: the UE searches neighbors at integer-subcarrier offsets around its serving lock.

\subsection{Limitations and System-Level Work}
\label{sec:limits}
Both simulations are snapshots; the residual and the delay change along a real pass. The two-cell simulation has isolated cells, one-subframe RSRP and exact UE pre-compensation.

\section{Conclusion}
\label{sec:conclusion}
Direct-to-cell service puts a satellite link on phones built for terrestrial networks, where a phone measures its neighbors at the frequency of its serving cell. In a satellite network, each cell arrives with its own residual Doppler, so a neighbor measured at the serving frequency reads low in RSRP or is not identified at all. The handover then comes late or not at all: in our field measurements, most cell changes end in radio-link failure, although a good neighbor had been measured. Link-level simulations reproduce this behavior: as the serving residual grows, the neighbor is first read low, then lost, and the handover report is never sent. Pre-compensating each cell’s residual at the UE restores the RSRP of the neighbor and its report. What the phone lacks today is the reference point of each cell and the location of its satellite, which no current signaling delivers.

\bibliographystyle{IEEEtran}
\bibliography{main}

\end{document}